\documentclass[prl,aps,twocolumn,showpacs,floatfix,superscriptaddress]{revtex4}
\usepackage{graphicx}
\usepackage{amsmath}
\usepackage{amssymb}
\usepackage{color}

\begin{document}

\title{Conductance of a helical edge liquid coupled to a magnetic
  impurity}

\author{Yoichi Tanaka}

\affiliation{Condensed Matter Theory Laboratory, RIKEN, Saitama 351-0198,
  Japan} 

\author{A. Furusaki}

\affiliation{Condensed Matter Theory Laboratory, RIKEN, Saitama 351-0198,
  Japan} 

\author{K. A. Matveev} 

\affiliation{Materials Science Division, Argonne National Laboratory,
  Argonne, Illinois 60439, USA}

\date{March 30, 2011}

\begin{abstract}

  Transport in an ideal two-dimensional quantum spin Hall device is dominated
  by the counterpropagating edge states of electrons with opposite spins,
  giving the universal value of the conductance, $2e^2/h$.
  We study the effect on the conductance of a magnetic impurity, which can
  backscatter an electron from one edge state to the other.  In the case of
  isotropic Kondo exchange we find that the correction to the electrical
  conductance caused by such an impurity vanishes in the dc limit,
  while the thermal conductance does acquire a finite correction due to
  the spin-flip backscattering.

\end{abstract}


\maketitle

Topological insulators have been actively studied in the last few
years \cite{review}.  In these systems the bulk of the sample is
insulating, with gapless electronic excitations allowed only at the
boundary.  One example of this phenomenon, the quantum Hall effect, is
realized in two-dimensional electron systems placed in a strong
magnetic field perpendicular to the sample.  Another example is the
so-called quantum spin Hall (QSH) effect, recently observed in HgTe quantum
wells \cite{konig}.  It is present in time-reversal invariant
two-dimensional systems, and can be viewed as two coexisting quantum
Hall states for opposite spin subsystems.  Although no external
magnetic field is present, an effective field is generated via
spin-orbit coupling.  The sign of the field is opposite for
spin-$\uparrow$ and spin-$\downarrow$ electrons, resulting in two edge
states propagating in opposite directions.

In the absence of defects, the conductance measured between two
contacts attached to the edges of a quantum spin Hall device is
$2e^2/h$.  (Here $e$ is the electron charge and $h$ is the Planck's
constant.)  This can be understood, e.g., as twice the conductance of
a standard quantum Hall structure, as the number of the edge states is
doubled.  The presence of two counterpropagating edge states also
allows for the possibility of backscattering of electrons, which would
reduce the conductance.
However, such backscattering processes are strongly restricted by the
time-reversal symmetry \cite{wu,xu}.  In particular, an impurity
without internal degrees of freedom cannot backscatter a single
electron at the edge, though backscattering of two electrons is
allowed.  Another allowed process is backscattering of a single electron
by a magnetic impurity.

The effect of a magnetic impurity on the conductance of a QSH device
was discussed recently by Maciejko \emph{et al.}\ \cite{maciejko}.
They argued that at high temperature the backscattering by
magnetic impurity gives a small negative correction to the
conductance.  As the temperature $T$ is lowered, the Kondo effect enhances
the backscattering and the correction grows.  The trend then reverses
at $T$ of the order of the Kondo temperature, and at $T=0$ the
quantized conductance $2e^2/h$ is restored.  In this Letter we study
the correction $\delta G$ to the conductance as a function of
frequency $\omega$, focusing on the case of isotropic Kondo exchange and
neglecting the two-particle backscattering.  Our results for $\delta
G(\omega)$ agree with the predictions by Maciejko \emph{et
  al.}\ \cite{maciejko} when $\omega$ exceeds a certain relaxation
rate, whereas in the dc limit $\omega\to0$ we find $\delta G=0$.

Following Refs.~\cite{wu,xu,maciejko}, we combine the two edge states
into a single nonchiral Tomonaga-Luttinger liquid and write its
Hamiltonian in the standard bosonized form \cite{giamarchi}
\begin{equation}
  \label{eq:Luttinger}
  H_0=\frac{\hbar v}{2\pi}
      \int dx[K(\partial_x\theta)^2+K^{-1}(\partial_x\phi)^2].
\end{equation}
Here $v$ is the velocity of the edge states, and parameter $K$ accounts
for the interactions between the electrons at the edge, with the repulsive
interactions corresponding to $K<1$.  The bosonic fields $\theta(x)$ and
$\phi(x)$ satisfy the commutation relations
$[\phi(x), \partial_y\theta(y)]=i\pi\delta(x-y)$.  For simplicity, we
limit ourselves to the case of spin-$\frac12$ impurity, and write the Kondo
coupling to the edge states using the standard boson representation of the
electron operators $\psi_{\uparrow,\downarrow}= (2\pi\alpha)^{-1/2}
e^{-i(\theta\pm\phi)}$, where $\alpha$ is the short distance cutoff and we
assigned spins up and down to the right- and left-moving electrons,
respectively.  We allow for the possibility of anisotropic coupling and
write the spin-flip term as
\begin{equation}
  \label{eq:spin-flip}
  H_{\perp}=\frac{J_\perp}{2\pi\alpha}[e^{2i\phi(0)}S^{+}+ e^{-2i\phi(0)}S^{-}],
\end{equation}
where $S^{\pm}=S^x\pm iS^y$ and $S^i$ are spin operators.  In
addition, the $z$ component of the impurity spin is coupled to the
Tomonaga-Luttinger liquid via
\begin{equation}
  \label{eq:z-coupling}
  H_z=-\frac{J_z}{\pi}\partial_x\theta(0)S^z.
\end{equation}
For convenience, our Hamiltonian allows for uniaxial exchange anisotropy,
which disappears at $J_z=J_\perp$.  Up to a minor difference in notation,
the Hamiltonian (\ref{eq:Luttinger})--(\ref{eq:z-coupling}) coincides with
the expression used in Refs.~\cite{wu,maciejko}.

To discuss the conductance of the device, one has to add to the
Hamiltonian a term accounting for the applied bias.  The most
straightforward approach is to assign different chemical potentials 
$\pm eV/2$ to the right- and left-moving electrons.
Such a perturbation takes the form
\begin{equation}
  \label{eq:voltage}
  H_V=-\frac{eV}{2\pi}\int dx\, \partial_x\theta.
\end{equation}
One should note that assigning the same chemical potential to all the
electrons moving in the same direction is an approximation.  It assumes
that the backscattering by impurity is weak, $\delta G\ll e^2/h$, and that
$\omega\ll v/L$, where $L$ is the distance between the source and drain 
contacts.

The operator (\ref{eq:voltage}) commutes with $H_0$ and $H_z$, and its
only effect is to change the energy by $\pm eV$ each time an electron is
backscattered and the impurity spin is flipped by $H_\perp$.  The same
effect is achieved by assigning different energy values to the up and down
components of the impurity spin, i.e., by replacing (\ref{eq:voltage})
with an effective magnetic field term
\begin{equation}
  \label{eq:Voltage}
  H_V=-eVS^z.
\end{equation}
This replacement can be more formally justified by noticing that the
difference of the operators (\ref{eq:voltage}) and (\ref{eq:Voltage})
commutes with the Hamiltonian.

To simplify the subsequent calculations it is convenient to rescale the
bosonic fields $\{\phi,\theta\}\to \{\sqrt{K}\phi, \theta/\sqrt{K}\}$ and
then perform the unitary transformation $U=e^{i\lambda\phi(0)S^z}$.  For
$\lambda=J_z/\pi\hbar v\sqrt K$ the additional term arising from the
transformation of $H_0$ cancels the coupling (\ref{eq:z-coupling}) of the
$z$ components of spins, resulting in the Hamiltonian
\begin{eqnarray}
  \label{eq:tildeH0}
  {\tilde H}_0&=&\frac{\hbar v}{2\pi}
        \int dx[(\partial_x\theta)^2+(\partial_x\phi)^2],
\\
  {\tilde H}_{\perp}&=&\frac{J_\perp}{2\pi\alpha}
         [e^{i2\sqrt{\tilde K}\phi(0)}S^{+}+ e^{-i2\sqrt{\tilde
             K}\phi(0)}S^{-}],
\label{eq:tildeHperp}
\end{eqnarray}
with the bias contribution (\ref{eq:Voltage}) retaining its form.  The
advantage of this procedure is that the effect of electron-electron
interactions and coupling of $z$ components of spins are now accounted for
by a single parameter $\tilde K=K(1-J_z/2\pi\hbar vK)^2$.

Our first goal is to evaluate the correction $\delta G(\omega)$ to the
conductance of the system due to the backscattering of electrons by the
impurity.  Since backscattering of a right-moving electron is always
accompanied by the impurity spin flip from $\downarrow$ to $\uparrow$, the
correction to the current can be computed as the time derivative of $S^z$,
i.e., $\delta I=-e\partial_t S^z$.  With the Hamiltonian in the form
(\ref{eq:Voltage})--(\ref{eq:tildeHperp}) one immediately obtains
\begin{equation}
  \label{eq:current_operator}
  \delta I = \frac{ie}{\hbar}\,\frac{J_\perp}{2\pi\alpha}
             [e^{i2\sqrt{\tilde K}\phi(0)}S^{+}- e^{-i2\sqrt{\tilde
             K}\phi(0)}S^{-}].
\end{equation}
In the linear response theory the conductance can be found using the Kubo
formula, which expresses $\delta G(\omega)$ in terms of the
current-current correlator.  The latter cannot be found exactly for
arbitrary values of the parameters $\tilde K$ and $J_\perp$.  Assuming weak
coupling (\ref{eq:tildeHperp}), one can evaluate $\delta G$ to the lowest
order in $J_\perp$, which results in the following expression:
\begin{eqnarray}
  \label{eq:conductance_Kubo}
  \delta G(\omega)&=&-\frac{2e^2}{\hbar^3} 
                     \left(\frac{J_\perp}{2\pi\alpha}\right)^2
                     \left(\frac{\pi T}{D}\right)^{2\tilde K}
                     \sin(\pi\tilde K)
\nonumber\\
  &&\times
  \frac{1}{i\omega}
  \int_0^\infty\frac{(e^{i\omega t}-1)dt}{[\sinh(\pi Tt/\hbar)]^{2\tilde K}}.
\end{eqnarray}
Here we have introduced the bandwidth $D=\hbar v/\alpha$.

In the most interesting regime of low frequencies, $\hbar\omega\ll T$, the
conductance is
\begin{equation}
  \label{eq:conductance_Kubo_result}
  \delta G=-\frac{e^2\gamma_0}{2T},
\end{equation}
where $\gamma_0$, defined as 
\begin{equation}
  \label{eq:gamma_0}
  \gamma_0=J_\perp^2\Upsilon,
\quad
  \Upsilon=\frac{[\Gamma(\tilde K)]^2}
                   {(2\pi\hbar)^2\alpha v\Gamma(2\tilde K)}
              \left(\frac{2\pi T}{D}\right)^{2\tilde K -1},
\end{equation}
has the meaning of the rate of spin-flip processes at $V=0$.  The same
result was obtained by a similar method by Maciejko
\emph{et al.}~\cite{maciejko}.
It is important to realize that the derivation relied on
the perturbation theory in $J_\perp$.  Let us now show that at low
frequencies $\omega\ll\gamma_0\propto J_\perp^2$ the conductance deviates
from (\ref{eq:conductance_Kubo_result}).

To this end we consider the dynamics of the impurity spin at finite
temperature and weak spin-flip coupling, $\gamma_0\ll T/\hbar$.  In this
case, the spin remains in the $\uparrow$ or $\downarrow$ state for a long
time. The rates of spin-flip events are easily computed and
take the form
\begin{equation}
  \label{eq:rates}
    \gamma_\pm=\gamma_0 \left(1 \pm \frac{eV}{2T}\right)
\end{equation}
for small voltage $eV\ll T$, where $\gamma_+$ and $\gamma_-$ are the rates
of up-flip and down-flip, respectively.

At any given moment the impurity spin can be in either $\uparrow$ or
$\downarrow$ state, and its behavior is described by the probabilities
$P_\uparrow$ and $P_\downarrow$.  Their time dependence can be found from
a simple rate equation
\begin{equation}
  \label{eq:rate-equation}
  \partial_t P_\uparrow=\gamma_+P_\downarrow-\gamma_- P_\uparrow
\end{equation}
and the condition $P_\uparrow+P_\downarrow=1$.  Each spin flip is
accompanied by backscattering of a single electron.  The resulting
correction to the electric current is $\delta I =
-e \partial_t P_\uparrow$.  To find the linear conductance one substitutes
the rates in the form (\ref{eq:rates}) with $V=V_0 e^{-i\omega t}$ into
(\ref{eq:rate-equation}), and calculates $\delta G=\delta
I/(V_0e^{-i\omega t})$.  This procedure yields
\begin{equation}
  \label{eq:deltaG}
  \delta G(\omega)=
        -\frac{e^2\gamma_0}{2T}\,\frac{\omega}{\omega+2i\gamma_0}.
\end{equation}
The rate equation approach is applicable only at relatively low
frequencies, $\hbar\omega\ll T$.  In a broad frequency range
$\gamma_0\ll\omega\ll T/\hbar$ expression (\ref{eq:deltaG}) recovers the
perturbative result (\ref{eq:conductance_Kubo_result}).

The results (\ref{eq:conductance_Kubo_result}) and (\ref{eq:deltaG})
differ dramatically in the dc limit, $\omega\to0$, where the correction
(\ref{eq:deltaG}) vanishes.  This is easily understood if one notices that
every time an electron is backscattered to the left (right) the impurity
spin is flipped up (down).  Thus, the backscattering current changes its
direction with every spin flip, and in the dc limit the correction to the
conductance vanishes.  This argument also applies at
$T\lesssim\hbar\gamma_0$, when the rate equation approach is no longer
applicable.

To illustrate this point, we consider a special case $\tilde K=\frac12$,
in which the dimension of the operator $\exp[i2\sqrt{\tilde K}\phi(0)]$ in
the definition of the spin-flip operator (\ref{eq:tildeHperp}) is
$\frac12$, \cite{maciejko}.  This case corresponds to the well-understood
Toulouse limit of the Kondo model, in which the Hamiltonian can be reduced
to that of a system of noninteracting chiral spinless fermions $\Psi(x)$
coupled to a discrete level:
\begin{eqnarray}
  \label{eq:Toulouse}
  H&=&-i\hbar v \int  \Psi^\dagger(x)\partial_x\Psi(x) dx
      -eV d^\dagger d
\nonumber\\
   && +\frac{J_\perp}{\sqrt{2\pi\alpha}}[d^\dagger\Psi(0)+\Psi^\dagger(0)d].
\end{eqnarray}
Here $d$ is the annihilation operator of a fermion at the discrete level
which models the impurity spin, $S^z=d^\dagger d-\frac12$.  The model
(\ref{eq:Toulouse}) is easily solved exactly, resulting in
\begin{equation}
  \label{eq:deltaG-Toulouse}
 \delta G(\omega)=-i
      \iint
      \frac{\omega[n(\xi_2)-n(\xi_1)]}{\hbar\omega-\xi_1+\xi_2+i\delta} 
      \frac{(e\Gamma/\pi)^2d\xi_1 d\xi_2}
           {(\xi_1^2+\Gamma^2)(\xi_2^2+\Gamma^2)}.
\end{equation}
Here the level width $\Gamma=J_\perp^2/4\pi\hbar\alpha v$ coincides with
$\hbar\gamma_0$ for $\tilde K=1/2$, Eq.~(\ref{eq:gamma_0}), and
$n(\xi)=1/(e^{\xi/T}+1)$ is the occupation probability of a fermion
state with energy $\xi$ measured from the Fermi level.  

\begin{figure}
\begin{center}
\includegraphics[scale=.8]{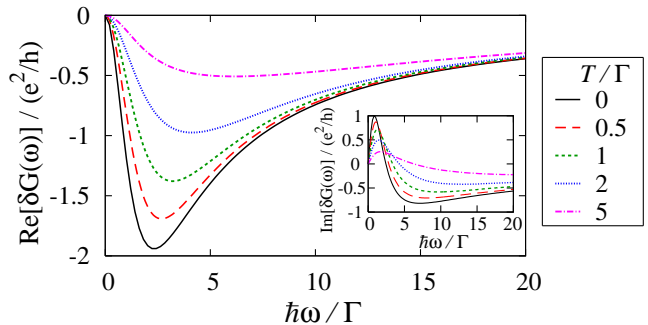}
\end{center}
\vspace{-5mm}
\caption{
(color online) 
Real part and imaginary part of 
$\delta G(\omega)$ for several temperatures $T/\Gamma$. 
}
\label{conReIm}
\end{figure}

In the high-temperature limit $T\gg\Gamma$ one can expand the difference
of the Fermi functions in the numerator to linear order in $\xi_2-\xi_1$.
This approximation recovers our earlier result (\ref{eq:deltaG}) with
$\gamma_0=\Gamma/\hbar$.  Importantly, the Toulouse limit expression
(\ref{eq:deltaG-Toulouse}) for $\delta G(\omega)$ does not rely on the
rate equation approach and is therefore valid for any temperature.  It is
easy to see that in the dc limit the expression (\ref{eq:deltaG-Toulouse})
vanishes for any $T$.  The dependence $\delta G(\omega)$ at various
temperatures is shown in Fig.~\ref{conReIm}.

In the absence of corrections to the quantized dc conductance of the
system, it is interesting to explore whether other transport
properties are affected by coupling to the impurity.  One such
property is the thermal conductance $\mathcal{K}$.  Let us assume that
the left- and right-moving electrons originate in the leads with
different temperatures, $T$ and $T+\Delta T$, respectively.  Spin-flip
scattering by the impurity gives rise to two kinds of electron
scattering processes.  The electrons are either scattered from the
warmer right-moving system to a colder left-moving one, or in the
opposite direction.  These two types of processes change the charge of
the right-moving system by $e$ and $-e$, respectively, resulting in no
correction to the time-averaged electric current.  On the other hand,
each scattering process transfers heat in the same direction: from the
warm right-moving system to the cold left-moving one.  As a result,
one expects to find a finite negative correction $\delta \mathcal K$
to the thermal conductance of the system.

To evaluate $\delta \mathcal K$, we identify the operator of the
difference of energies in the right- and left-moving branches
$\delta\tilde H=(\tilde H_0^R-\tilde H_0^L)/2$.  The energy densities of
the two subsystems are given by $\hbar
v[\partial_x(\phi\mp\theta)]^2/4\pi$, leading to
\begin{equation}
  \label{eq:energy_difference}
  \delta\tilde H = 
    -\frac{\hbar v}{2\pi}\int\partial_x\phi\,\partial_x\theta dx.
\end{equation}
The operator  $\delta J_E$ of the energy current
transferred between the right- and left-moving branches is then found as
the time derivative of (\ref{eq:energy_difference}), resulting in
\begin{equation}
  \label{eq:energy_current}
  \delta J_E = -\frac{\hbar v\sqrt{\tilde K}}{e}\,
               \partial_x\phi(0)\,
               \delta I,
\end{equation}
where $\delta I$ is the electric current operator in Eq.\ 
(\ref{eq:current_operator}).

It is worth noting that the operators $\partial_x\phi(0)$ and $\delta I$
act in separate subspaces of odd and even $\phi(x)$.  In
particular, the dynamics of $\partial_x\phi(0)$ is not affected by the
coupling (\ref{eq:tildeHperp}) to the impurity.  A similar observation was
made by Kane and Fisher \cite{kanefisher} in the study of thermal
transport of a Luttinger liquid through a tunneling barrier.  They used
the factorization of the energy current operator similar to
Eq.~(\ref{eq:energy_current}) to obtain a relation between the thermal and
electrical conductances of the system.  Repeating their procedure for our
Hamiltonian, we obtain the expression
\begin{equation}
  \label{eq:Kane_relation}
  \delta{\mathcal K} = \frac{\hbar^3\tilde K}{8e^2T^2}
           \int_{-\infty}^\infty \frac{\omega^2{\rm Re\,}\delta G(\omega)}
                              {\sinh^2(\hbar\omega/2T)}d\omega,
\end{equation}
fully analogous to Eq.~(19) of Ref.~\cite{kanefisher}.

The relation (\ref{eq:Kane_relation}) shows that thermal conductance is
determined by the electrical conductance at frequencies $\omega \sim
T/\hbar$.  As a result, the fact that $\delta G$ vanishes at $\omega\to0$
does not mean that there will be no correction to $\mathcal K$.  For
instance, in the lowest order in spin-flip scattering
(\ref{eq:tildeHperp}), one can find $\delta \mathcal K$ by substituting
into Eq.~(\ref{eq:Kane_relation}) $\delta G(\omega)$ in the form
(\ref{eq:conductance_Kubo}).  This procedure yields
\begin{equation}
  \label{eq:deltaKperturbative}
  \delta \mathcal K = -\frac{\pi^2}{2}\, 
                       \frac{\tilde K^2}{2\tilde K +1}\,\gamma_0.
\end{equation}
Within the applicability of the perturbation theory, $\hbar\gamma_0\ll T$,
this correction is small compared to the nominal value ${\mathcal
  K}_0=\pi^2T/3h$ of the thermal conductance of a Kramers' pair of edge
states.

At $\hbar\gamma_0\sim T$ the perturbative approach is not applicable, but
an exact solution is possible in the Toulouse limit, $\tilde K=1/2$.
Substituting the conductance (\ref{eq:deltaG-Toulouse}) into
Eq.~(\ref{eq:Kane_relation}) we obtain
\begin{equation}
  \label{eq:deltaK-Toulouse}
   \delta \mathcal K = \frac{\Gamma^2}{8h T^2}
        \iint\frac{(\xi_2-\xi_1)^3[n(\xi_2)-n(\xi_1)]d\xi_1d\xi_2}
        {(\xi_1^2+\Gamma^2)(\xi_2^2+\Gamma^2)
         \sinh^2[(\xi_2-\xi_1)/2T]}.
\end{equation}
At high temperatures ($T\gg \Gamma$),
$\delta\mathcal{K}=-\pi^2\Gamma/16\hbar$,
in agreement with the perturbative result
(\ref{eq:deltaKperturbative}).  On the other hand, the fact that ${\rm
  Re}\,\delta G(\omega)\propto\omega^2$ at $\omega\to0$ means that the
correction $\delta\mathcal{K}$ in Eq.\ (\ref{eq:Kane_relation})
is suppressed as $T^3$ at low temperatures, $T\ll \Gamma$.  Indeed, from
Eq.~(\ref{eq:deltaK-Toulouse}) one finds $\delta\mathcal
K=-2\pi^3T^3/15\hbar\Gamma^2$ in this regime.

\begin{figure}
\begin{center}
\includegraphics[scale=.8]{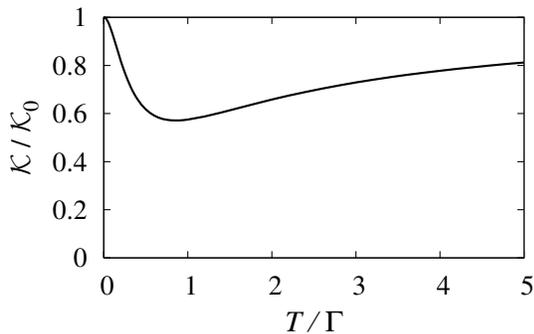}
\end{center}
\vspace{-5mm}
\caption{Nonmonotonic temperature dependence of the thermal
  conductance $\mathcal K=\mathcal K_0+\delta\mathcal K$ in the
  Toulouse limit, Eq.~(\ref{eq:deltaK-Toulouse}).
\label{fig:non-monotonic}}
\end{figure}

The suppression of $\delta\mathcal K$ in both the low- and
high-temperature limits results in nonmonotonic behavior of the
normalized thermal conductance $\mathcal K/\mathcal K_0$ as a function of
temperature, Fig.~\ref{fig:non-monotonic}.  Such nonmonotonic behavior
was predicted for the electrical conductance by Maciejko \emph{et al.}
\cite{maciejko}, based on the well-known nonmonotonic temperature
dependence of the spin-flip scattering in the Kondo problem.  Although our
theory predicts quantized dc conductance $2e^2/h$ at any temperature, the
nonmonotonic temperature dependence is recovered for thermal transport.

At $T\lesssim\gamma_0$ and $\omega\sim\gamma_0$ our correction to the
electrical conductance can take rather large values $|\delta G(\omega)|
\sim e^2/h$, see, e.g., Fig.~\ref{conReIm}.  We do not believe the
correction $\delta G(\omega)$ has a clear physical meaning in this regime,
as the conditions for the applicability of the model (\ref{eq:voltage})
are violated.  This caveat, however, does not apply to our main
conclusion, namely vanishing $\delta G$ at $\omega\to0$.  Similarly, the
behavior of the thermal conductance shown in Fig.~\ref{fig:non-monotonic} is
only qualitatively correct at $T\sim\Gamma$, where the relative correction
$\delta\mathcal K/\mathcal K_0$ is of order unity.

The main feature of our work, the absence of correction to the dc
conductance due to Kondo scattering, is caused by the symmetry of the
Hamiltonian (\ref{eq:Luttinger})--(\ref{eq:z-coupling}) that preserves
the $z$ component of the total spin of the system.  Our model is
justified if the magnetic impurity is approximated as a single-orbital
Anderson impurity, in which case the Kondo exchange is isotropic,
$J{\boldsymbol\sigma}\cdot{\boldsymbol S}$, even in the presence of
spin-orbit coupling \cite{shekhtman}.  Anisotropic corrections may
appear in the case of multiple orbitals \cite{anisotropicexchange}.
Exchange anisotropy can break the conservation of the $z$ component of
the total spin and result in nonvanishing correction to the dc
conductance.  In the simplest model of anisotropic exchange
$J_i\sigma^iS^i$ with $J_x\neq J_y$ the correction to the dc
conductance above the Kondo temperature is easily found from the rate
equation approach,
\begin{equation}
  \label{eq:anisotropic_conductance}
  \delta G = -\frac{e^2\Upsilon}{4T}\,\frac{(J_x^2-J_y^2)^2}{J_x^2+J_y^2},
\end{equation}
with $\Upsilon$ defined by Eq.~(\ref{eq:gamma_0}).  At $T$ below the Kondo
temperature the impurity spin is screened, and the effect of exchange
anisotropy is reduced to that of an impurity-induced two-electron
backscattering.  The effect of these processes on conductance was studied in
Ref.~\cite{maciejko} and found to be suppressed at $T\to0$ as $T^{2(4K-1)}$.

Another limitation of our model is the assumption that all electrons
moving in the same direction have the same spin.  Kramers degeneracy
guarantees that electrons with momenta $\pm p$ have opposite spins,
and we defined the $z$ axis as the direction of the spin of the
right mover at the Fermi level, which is determined by the specific
form of spin-orbit coupling.
Transport is controlled by electrons
with momenta $|p\pm p_F |\sim T/v$, whose spins may deviate slightly
from the $z$ direction defined at $p=p_F$.  This deviation may result
in a correction to the dc conductance vanishing at low temperatures as
a power of $T/D$.

The authors are grateful to B. I. Halperin and S. C. Zhang for helpful
discussions.
A.F. and K.A.M. are grateful to the Aspen Center for Physics, where
part of the work was performed, for hospitality.  This work was
supported by a Grant-in-Aid for Scientific Research from JSPS, Japan
(No.~21540332) and by the U.S. DOE, Office of Science, under Contract
No.~DE-AC02-06CH11357.

\end{document}